\documentclass[fleqn,usenatbib]{mnras}

\usepackage{newtxtext,newtxmath}
\usepackage{subfigure}
\usepackage[T1]{fontenc}
\usepackage{ae,aecompl}

\usepackage{graphicx}	% Including figure files
\usepackage{amsmath}	% Advanced maths commands
\usepackage{amssymb}	% Extra maths symbols
\usepackage{color}
\definecolor{Red}{rgb}{0.9,0,0}
\definecolor{Blue}{rgb}{0,0,0.9}

\title[Mutual approximations of the Galilean moons]{APPROX -- Mutual approximations between the Galilean moons. The 2016-2018 observational campaign.}

\author[B. Morgado et al.]{B. Morgado$^{1,2}$\thanks{E-mail:Morgado.fis@gmail.com},
R. Vieira-Martins$^{1,2,3}$,  
M. Assafin$^{2,3}$,
D. I. Machado$^{4,5}$, \newauthor
J. I. B. Camargo$^{1,2}$,   
R. Sfair$^{6}$, 
M. Malacarne$^{7}$, 
F. Braga-Ribas$^{1,2,8}$, 
V. Robert$^{9,10}$,  \newauthor
T. Bassallo$^{1}$,
G. Benedetti-Rossi$^{1,2}$, 
L. A. Boldrin$^{6}$,  
G. Borderes-Motta$^{6}$, \newauthor 
B. C. B. Camargo$^{6}$, 
A. Crispim$^{8}$, 
A. Dias-Oliveira$^{1,11}$, 
A. R. Gomes-J\'unior$^{2,3,6}$,\newauthor
V. Lainey$^{12,10}$,
J. O. Miranda$^{7}$, 
T. S. Moura$^{6}$,
F. K. Ribeiro$^{7}$, 
T. de Santana$^{6}$,\newauthor 
S. Santos-Filho$^{2}$,
L. L. Trabuco$^{5}$, 
O. C. Winter$^{6}$ 
and T. A. R. Yamashita$^{6}$  
\\
$^{1}$Observat\'orio Nacional/MCTIC, R. General Jos\'e Cristino 77, Rio de Janeiro, RJ 20.921-400, Brazil\\
$^{2}$Laborat\'orio Interinstitucional de e-Astronomia - LIneA, Rua Gal. Jos\'e Cristino 77, Rio de Janeiro, RJ 20921-400, Brazil\\
$^{3}$Observat\'orio do Valongo/UFRJ, Ladeira Pedro Antonio 43, Rio de Janeiro, RJ 20080-090, Brazil\\
$^{4}$Universidade Estadual do Oeste do Paran\'a (Unioeste), Avenida Tarqu\'inio Joslin dos Santos 1300, Foz do Igua\c{c}u, PR 85870-650, Brazil\\
$^{5}$Polo Astron\^omico Casimiro Montenegro Filho/FPTI-BR, Avenida Tancredo Neves 6731, Foz do Igua\c{c}u, PR 85867-900, Brazil\\
$^{6}$UNESP - S\~ao Paulo State University, Grupo de Din\^amica Orbital e Planetologia, CEP 12516-410, Guaratinguet\'a, SP 12516-410, Brazil\\
$^{7}$Universidade Federal do Esp\'irito Santo, Av. Fernando Ferrari 514, Vit\'oria, ES 29075-910, Brasil\\
$^{8}$Federal University of Technology - Paraná (UTFPR/DAFIS), Av. Sete de Setembro, 3165, CEP 80230-901 - Curitiba - PR - Brazil\\
$^{9}$Institut Polytechnique des Sciences Avanc\'ees IPSA, 63 bis Boulevard de Brandebourg, 94200 Ivry-sur-Seine, France\\
$^{10}$IMCCE, Observatoire de Paris, PSL Research University, CNRS-UMR 8028, Sorbonne Universit\'es, UPMC, Univ. Lille 1, 77 \\Av. Denfert-Rochereau, 75014 Paris, France\\
$^{11}$Escola SESC de Ensino M\'edio, Avenida Ayrton Senna, 5677, Rio de Janeiro - RJ, 22775-004, Brazil\\
$^{12}$Jet Propulsion Laboratory, California Institute of Technology, 4800 Oak Grove Drive, Pasadena, CA 91109-8099, United States}

\date{Accepted XXX. Received YYY; in original form ZZZ}

\pubyear{2018}

\begin{document}
\label{firstpage}
\pagerange{\pageref{firstpage}--\pageref{lastpage}}
\maketitle

\begin{abstract} %less than 250 words
The technique of mutual approximations accurately gives the central instant at the maximum apparent approximation of two moving natural satellites in the sky plane. This can be used in ephemeris fitting to infer the relative positions between satellites with high precision. Only the mutual phenomena -- occultations and eclipses -- may achieve better results. However, mutual phenomena only occur every six years in the case of Jupiter. Mutual approximations do not have this restriction and can be observed at any time along the year as long as the satellites are visible. In this work, we present 104 central instants determined from the observations of 66 mutual approximations between the Galilean moons carried out at different sites in Brazil and France during the period 2016--2018. For 28 events we have at least two independent observations. All telescopes were equipped with a narrow-band filter centred at 889 nm with a width of 15 nm to eliminate the scattered light from Jupiter. The telescope apertures ranged between 25--120 cm. For comparison, the precision of the positions obtained with classical CCD astrometry is about 100 mas, for mutual phenomena it can achieve 10 mas or less and the average internal precision obtained with mutual approximations was 11.3 mas. This new kind of simple, yet accurate observations can significantly improve the orbits and ephemeris of Galilean satellites and thus be very useful for the planning of future space missions aiming at the Jovian system.
\end{abstract}

\begin{keywords}
Methods: data analysis -- Astrometry -- Planets and satellites: individual: Io, Europa, Ganymede, Callisto.
\end{keywords}

%%%%%%%%%%%%%%%%%%%%%%%%%%%%%%%%%%%%%%%%%%%%%%%%%%

%%%%%%%%%%%%%%%%% BODY OF PAPER %%%%%%%%%%%%%%%%%%

\section{Introduction} \label{intro}

The orbital studies of the natural satellites can give us hints about the formation processes of these moons \citep{Charnoz2011,Crida_2012}. Also, they can give us valuable information about their interiors, with accurate estimations of the tidal effect. One example is the thermal equilibrium in Io (the innermost of the Galilean moons), determined by the agreement between the orbital energy loss and the heat evacuated at Io's surface  \citep{Lainey2009}.

Improvement of the orbits on these studies demand systematic astrometry of these moons, preferably over extended periods of time and, as much as possible, with accurate and precise measurements. These measurements, observables in a more general sense, are fitted with the use of dynamical models \citep{Sitter1928,Lieske1987,Lainey2009}. For instance, improvements in the study of the tidal force in the Jovian system requires positions with a precision better than 30 mas (milliarcseconds) \citep{Lainey2016}.

Usual CCD astrometry relies on the imaging of the target in the Field of View (FOV) with an adequate number of cataloged reference stars. For the Galilean moons, this is not an easy task. Jupiter brightness (Magnitude in V band around $-$2.5) makes it difficult to image cataloged stars (V = 12 to 20), since Jupiter saturates and spreads its light all over the FOV with longer exposures. Methods to reduce this brightness have been tried, however the precision in a classical CCD astrometry of a single satellite is yet not satisfactory, i.e. the standard deviation of ephemeris residuals from a few hundred observations per night ranges between 100-150 mas \citep{Kiseleva2008}.

Mutual occultations and eclipses furnish very precise relative positions between two satellites. The drawback is that they can only be observed during the equinox of the host planet, when the Earth and the Sun pass through the orbital plane of the satellites. In the case of Jupiter it happens every 6 years, for Saturn every 15 and for Uranus every 42 \citep{Arlot2012,Arlot2013,Arlot2014}. For the Galileans satellites, mutual phenomena may deliver relative positions with a precision better than 5 mas \citep{Diasoliveira2013,Emelyanov2009}. More than 600 light curves were obtained in the last Mutual phenomena campaign between the Galilean moons, the PHEMU15, with an average precision of 24 mas \citep{Saquet2018}.

This scenario motivated the search for alternative methods to furnish astrometric data for these satellites. For example, \cite{Peng2012} determined relative positions between a pair of satellites when they are close together in the FOV, with a relative distance smaller than 85 arcseconds, and obtained precisions of 30 mas in these relative positions.

A more recent attempt is the mutual approximations technique developed by us \citep{Morgado2016}, primarily suggested by \cite{Arlot1982}. In this method, the instant of the maximum apparent approximation in the sky plane between two moving satellites can be determined with a precision that corresponds to less than 10 mas. The technique is immune to two problems in the CCD astrometry of the Galilean moons: the determination of the pixel scale and the orientation of the CCD with respect to the right ascension and declination axes in the sky \citep{Emelyanov2017}. Also, the observations with this technique are easily done with small aperture size telescopes (few centimeters). One important aspect of the method is the correct registering of time. Fortunately, this is also usually easy to accomplish with GPS receivers, specialized software or internet services that calibrate the acquisition computer's UTC time inserted in the images.

In this paper we detail the APPROX, an observational campaign of mutual approximations between the Galilean moons. It is a collaboration between Brazilian and French institutes, with six observational sites. This campaign observed 66 mutual approximations, obtaining 104 distance curves between February 2016 and August 2018. The average precision of the central instant was 11.4 mas using the relative velocity in each event to convert between seconds of time and arcseconds. We also present a procedure to use the mutual approximation data as observables to determine the parameters of the satellites' orbits in ephemeris fitting.

Section \ref{Method} brings an overview of the mutual approximation method. In Section \ref{Observacao}, we describe the observational campaign, prediction, simulations, observations and how we processed the observed data. In Section \ref{res_APPROX}, we present our results. We describe in Section \ref{ajuste} a procedure to use the central instants of mutual approximations on ephemeris fitting. We set our conclusions in Section \ref{conclusao}.

\section{THE METHOD OF MUTUAL APPROXIMATIONS} \label{Method}

A detailed description of the method of mutual approximations is given in \cite{Morgado2016}. Here, we briefly summarize the principles of this method.

It is possible to determine the ICRS position of a target in a CCD frame but it is necessary an astrometric star catalog to find the pixel scale, the CCD orientation and the zero point. This last is not needed when fitting ephemeris of natural satellites if the relative distances are known.

However, in the case of the Galilean satellites, it is not trivial to determine the pixel scale and image orientation without catalog reference stars. One possibility is to use the measured relative satellite positions and motions and a reference ephemeris as template \citep{Peng2012}, but then the ``true'' relative distances may be masked by the correlation with the errors in the ephemeris scale and orientation.

In the mutual approximation technique, we do not work with scaled distances, but with instrumental ones given pixel units. Scaled distances can be derived, but are not used in any fitting for finding the central instant and impact parameter and their errors - only instrumental distances are used for that. We can derive the pixel scale and CCD orientation using an ephemeris as template, or we can use the nominal pixel scale of the instrument, but only for internal checking purposes, like converting the impact parameter and its uncertainties to mas in order to have a better evaluation of the fit. We emphasize that post-derived scaled distances between the satellites are not the primary result of the method of mutual approximations.

Anyway, the main result of the method is the central instant at maximum apparent approximation between two satellites. Thus, we must calibrate time correctly, preferably with a precision better than  0.1 seconds. We do that using GPS receivers or with the time-calibration software called Dimension$^4$ \footnote{Website:\url{http://www.thinkman.com/dimension4/}}.

The mutual approximation method consists on fitting the apparent distances $s_{ij}$ in the sky plane between two moving satellites $i$ and $j$ by a N-degree polynomial in time, defined by equation \eqref{eq_poly}. The underlying assumption - which actually defines a mutual approximation - is that $s_{ij}$ gradually decreases with time, reaches a minimum and then starts to increase. The use of the square of $s_{ij}$ simplifies computations. The degree of the polynomial is determined by evaluating tentative fittings to the (squared) apparent distances computed from a reference ephemeris.

\begin{equation}\label{eq_poly}
s_{ij}^2(t) = \sum_{n=1}^{N} a_{n}.t^{n}	
\end{equation}

We then use the fitted polynomial coefficients ($a_n$) to determine the central instant $t_0$ (the instant at maximum approximation, when the apparent distance is minimum), the impact parameter $d_0$ (the minimum apparent distance in the sky plane between both satellites, which occurs by definition at $t_0$ when the approximation is at maximum), the relative velocity $v_0$ at $t_0$ between both satellites in the sky-plane, and their uncertainties ($\sigma d_0$, $\sigma t_0$ and $\sigma v_0$). The central instant is always obtained in UTC. When fitting observations, the impact parameter and its uncertainty is obtained in pixels, and the relative velocity and its uncertainty in pixels per second of time. When fitting reference ephemeris data, the impact parameter is generally computed in arc seconds and the relative velocity in arcseconds per second of time.

We also correct a shift in the observed central instant due to effects on the apparent distances caused by: a) the different apparent sizes of each satellite and the solar phase angle correction; b) atmospheric refraction; c) diurnal and annual aberration. The correction is determined after comparing the shift in the central instant obtained from fittings using a reference ephemeris with and without these effects. The shift usually ranges from 1 to 6 seconds (5 to 30 mas). 

Marks in the surface, or topography, of the satellites could affect the centroid measurement. As pointed out by \cite{Lindegren1977}, the maximum offset could be 35 mas, and would affect systematically all astrometric measurements during a run. Only a very precise albedo map of these satellites in the spectral region of the observations (in our case 889 nm) would allow us for inferring exactly its contribution. We highlight, however, that for mutual approximations, we would be affected only by a fraction of this offset, along the direction of relative motion between both satellites.

After all the fittings and computation of all parameters, for analysis and comparison purposes, we only use non-squared apparent distances. Once ordered in time, we have the distance curves of the event. There are two kinds of observed distance curves -- measured and fitted -- and two kinds of ephemeris distance curves -- ephemeris- and fitted-based. The nature of the distance curves discussed along the text should be clear from the context.

\section{OBSERVATIONAL CAMPAIGN} \label{Observacao}

An observational campaign starts with the prediction of the apparent close approximations between two satellites that are really interesting. The second step is the simulation of these events, which can give some hints about the best instrumental configuration and observation procedures for the participants of the campaign. The third step is the observation itself and for last the analysis of the data acquired.     

\subsection{PREDICTION} \label{predicao}

The predictions of the approximations were made with the topocentric ephemeris for each participating observatory using the NAIF SPICE\footnote{Website: \url{ http://naif.jpl.nasa.gov/naif/}} toolkit, Jovian ephemeris \emph{jup310} and planetary ephemeris DE430. 

The \emph{precision premium} \citep{Peng2008} predicts an increase in the precision in the measurement of apparent distances between two objects in the sky plane when this distance is smaller than 85 arcseconds. In this scenario, we avoid the effects of distortions in the FOV, since both satellites should be affected in the same way. In order to avoid a prohibitive number of events, we only chose the approximations with a impact parameter smaller than 30 arcseconds.

We selected all the mutual approximations that were visible for the observatories with elevation above 30$^{o}$. We set a minimum apparent distance of 10 arcseconds between both satellites and the Jupiter limb. In total, we predicted 102 events between February 2016 and August 2018. From these, we observed 66 mutual approximations - the others were lost due to bad weather conditions or instrumental issues. 

\subsection{SIMULATIONS} \label{simulacao}

In order to test the feasibility of the selected events, it is important to simulate observations and analyze the different aspects that rise in each scenario. First, we study the expected precision of the mutual approximation's central instant ($\sigma t_{0}$) and impact parameter ($\sigma d_{0}$) for different values of time resolution ($\delta t$) and signal to noise ratio (SNR). Second, we evaluate the ideal duration of the observation of a mutual approximation. Finally, we mimic the presence of gaps in the distance curve, which are often caused by weather or instrumental issues.

Let's illustrate all these three steps in the simulation of the events by taking as example the approximation between Io (501) and Ganymede (503) that happened on February $24^{th}$, 2016. We added a Gaussian error with standard deviation equal to $\sigma_{noise}$ in the distance between the pair of satellites to simulate real observations. We repeated the simulation 100 times with normalization to remove random systematic errors.

In the first step we studied how the central instant and impact parameter errors are affected by different SNR and different $\delta t$ (the time difference between two consecutive images). It is clear that the best case scenario is a high SNR and a low $\delta t$. However, part of the time resolution is related with the time exposure, which in turn is correlated with the SNR. Thus, the simulations in this step show us which parameter we have to prioritize to obtain the best results. 

\begin{figure}
\includegraphics[height=07cm]{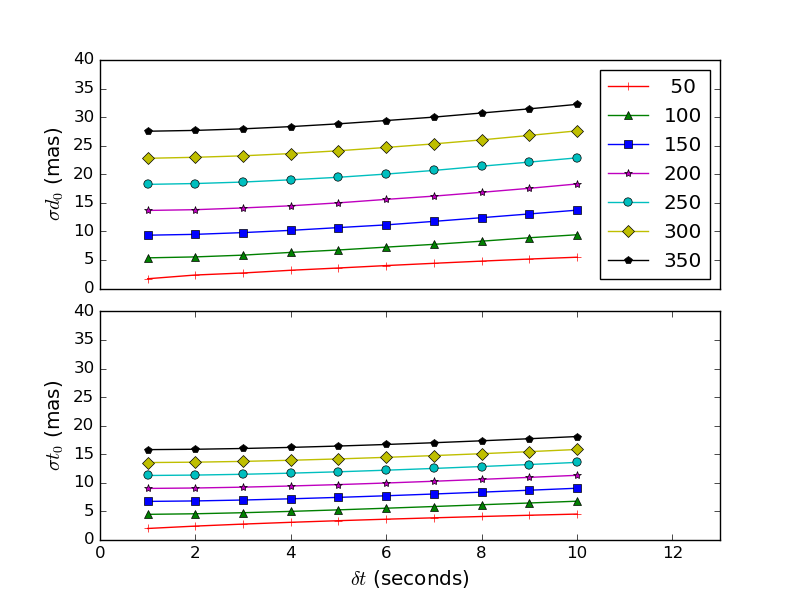}
\caption{Simulating the observation of a mutual approximation with different values for the Signal to Noise Ratio (SNR - $\sigma_{noise}$) and time resolution ($\delta t$); the x-axis is $\delta t$, the upper y-axis is the error of the impact parameter ($\sigma d_{0}$) and the bottom y-axis the error of the central instant ($\sigma t_{0}$), both in mas. The different colours and marks represent different $\sigma_{noise}$ regimes, in mas.}
\label{Fig:testeSNR}
\end{figure}

For the simulations in this step, we chose $\delta t$ ranging between 1.0 \-- 10.0 seconds and the $\sigma_{noise}$ between 50 \-- 350 mas. The result is displayed in Fig. \ref{Fig:testeSNR} where we can see that a high $\sigma_{noise}$ (low SNR) affects more the precision of the impact parameter ($\sigma d_{0}$) and the central instant ($\sigma t_{0}$) than the time resolution itself. This means that a good SNR should be prioritized in the observations. For the remaining of the simulations, we used $\delta t~=~4$ s and $\sigma_{noise}~=~100$ mas, which are the mean values in \cite{Morgado2016}.  

In the second step, the simulations has the purpose to evaluate the duration that a mutual approximation should be observed. We simulate observations starting one hour before and ending one hour after the central instant ($t_{0}$). We eliminated pairs of simulated images symmetrically placed around the central instant, one pair at a time, until our model failed to determine the central instant. This happens for $\Delta t$ smaller than 10 minutes (five minutes for each side around $t_{0}$).

\begin{figure}
\includegraphics[height=07cm]{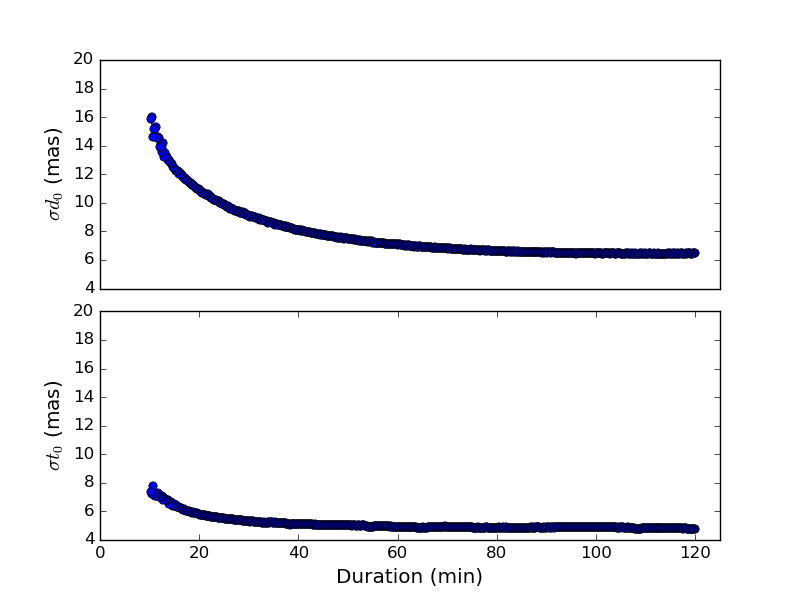}
\caption{Simulating the duration of the mutual approximation event; the x-axis is the duration of the event ($\Delta t$), symmetrical with regard to the central instant ($t_{0}$); the upper y-axis is the error of the impact parameter ($\sigma d_{0}$) and the bottom y-axis the error of the central instant ($\sigma t_{0}$), both in mas.}
\label{Fig:testeduracao}
\end{figure}               

As seen in Fig. \ref{Fig:testeduracao} for events with duration between 120 and 40 minutes there is no significant difference in the precision of the result obtained for the central instant. For the next simulations, we used distance curves of 60 minutes with 30 minutes before and after the central instant.

In the third step, the simulations mimic problems that arise from instrument issues and/or bad weather conditions, for evaluating how the absence of points along the curve affects the error of the results. The simulations are subdivided in two different scenarios: ($i$) gaps are present along all the curve; ($ii$) only one side of the curve is available. Both these scenarios were explored in \cite{Morgado2016}, but here we study them in detail.

\begin{figure}
\includegraphics[height=07cm]{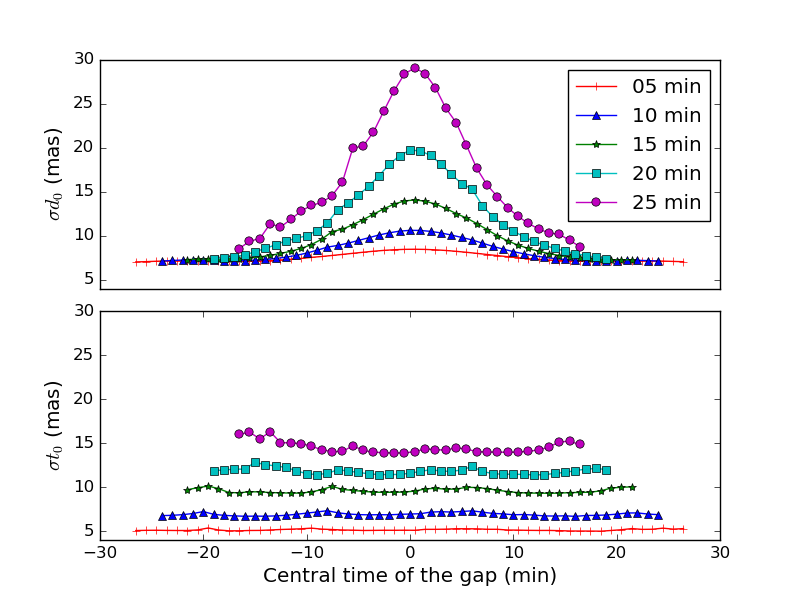}
\caption{Simulating gaps in a mutual approximation; the x-axis is the central time of the gap $t_{gap}$, the upper y-axis is the error of the impact parameter ($\sigma d_{0}$) and the bottom y-axis the error of the central instant ($\sigma t_{0}$), both in mas. The different colours and marks represent different sizes of the gap in minutes.}
\label{Fig:testegap}
\end{figure}

In scenario ($i$) not only the size of the gap matters, but also the gap location. We explored gaps with sizes $\Delta t_{gap}$ equal to 5, 10, 15, 20 and 25 minutes, in different positions along the distance curve with respect to the central time of the gap $t_{gap}$. In Fig. \ref{Fig:testegap}, we plot the errors of the impact parameter ($\sigma d_{0}$) and the central instant ($\sigma t_{0}$) over $t_{gap}$. We remark that the location of the gap does not affect $\sigma t_{0}$, however $\sigma d_{0}$ is strongly affected by gaps near the central instant. This kind of gap always occurs during the mutual phenomena period when the mutual approximation culminates in an occultation, during which of course it is not possible to measure the (x,y) centroids of both satellites individually. 

\begin{figure}
\includegraphics[height=07cm]{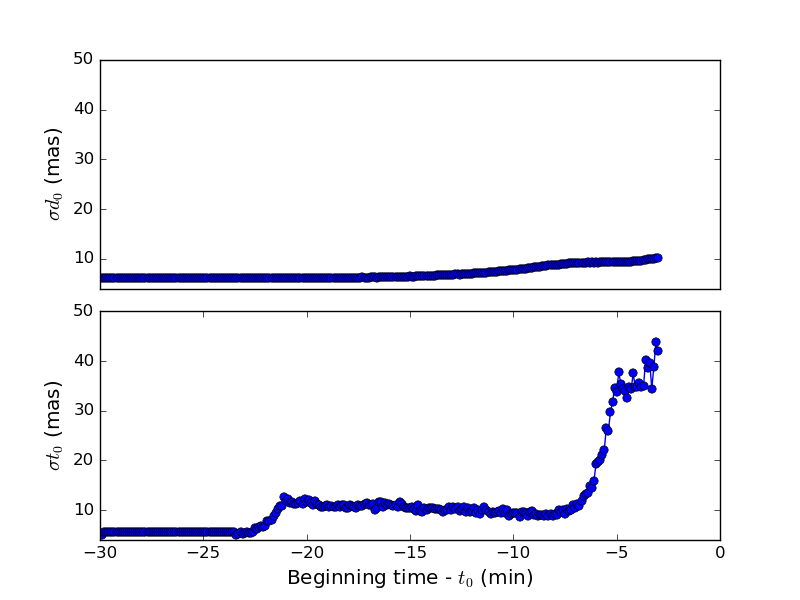}
\caption{Simulating one-sided curves in a mutual approximation; the x-axis is the beginning time minus the central instant in minutes; the upper y-axis is the error of the impact parameter ($\sigma d_{0}$) and the bottom y-axis the error of the central instant ($\sigma t_{0}$), both in mas.}
\label{Fig:testelado}
\end{figure}

The distance curve in a mutual approximation should naturally be a quasi-symmetrical one with respect to the central instant. Thus, observing only one side of the curve precludes a good determination of the central instant. In scenario ($ii$) we investigated how close to the central instant we can start (or finish) one observation and still get a good precision. In the simulations, one by one, we eliminate points only from one side of the curve and compute the errors in the impact parameter ($\sigma d_0$) and the central instant ($\sigma t_0$). The results can be seen in Fig. \ref{Fig:testelado}. It shows that the central instant error is strongly affected by the absence of only one side of the curve. For observations starting less than 5 minutes before the central instant, the error can achieve 30 mas or more.

The first and second steps of simulations show us the eventual limitations of the method for each event. They allow us for alerting the observers to take the necessary precautions in their instrumental setup and observational strategies. This optimizes the outputs of each event and ultimately improves the overall results of the campaign. The scenarios ($i$ and $ii$) in the third step of simulations show us what to expect in the realistic case of a mutual approximation event with weather and/or instrument issues. It is noteworthy that even in the worse scenarios with errors greater than 30 mas, the measurements may still be very useful in the ephemeris fitting of the Galilean satellites.

\subsection{OBSERVATIONS} \label{campanha}

The observations were made at five different sites in the South and South-East of Brazil and one site in the South-East of France. The geographical longitude, latitude, altitude and the  \emph{Minor Planet Center} (MPC) Observatory code of the sites (XXX for sites without a code) of each observatory are listed in Table \ref{tb:obs_sites}, which also displays instrumental information for each site, the observers and the number of positive detections. Note that the aperture diameters of the telescopes ranged between 25-120 cm. We also show .

We encouraged the coverage of each event by multiple sites in order to lose as few events as possible due to bad weather or instrumental problems, not to mention the advantage gained in the analysis. Also, we oriented the observers to place the satellites of the mutual approximation in the central part of the CCD FOV to attenuate the effects of field distortions, if any (see \cite{Peng2012b}).

All the observations were made with a narrow-band filter centred at 889 nm with a width of 15 nm. Radiation in this spectral range is absorbed by the methane in the Jupiter's high clouds, making its albedo drop to values below 0.1 \citep{Karkoschka1994, Karkoschka1998}. This filter is very efficient in decreasing the scattered light of Jupiter without affecting the brightness of the satellites, as pointed out by \cite{Karkoschka1994}. Thus, we could obtain good SNR images of the satellites (V around 5) with exposures of a few seconds without the interference of the scattered light of the planet.

\begin{table*}
\begin{center}
\caption{2016-2018 mutual approximations campaign. Observation information.}
\begin{tabular}{lcccc}
\hline
\hline
Site  & Longitude & Observers & Telescope aperture & No. of events \\
Alias & Latitude & Team&CCD&detected \\
MPC code& Altitude &&Pixel scale&         \\
\hline
\hline
Itajub\'a/MG-Brazil      &45$^o$ 34' 57.5" W& B. Morgado   & 60 cm   & 29 \\
OPD             &22$^o$ 32' 07.8" S& J. I. B. Camargo   & Andor/Ikon    &    \\
874             &1864 m            & T. Bassallo        & 0.37 "/px &  \\ 
                &                  & A. R. Gomes-J\'unior &         &    \\
                &                  & S. Santos-Filho    &         &    \\
				&                  & A. Dias-Oliveira   &         &    \\
                &                  & G. Benedetti-Rossi &         &    \\
\hline
Foz do Igua\c{c}u/PR-Brazil&54$^o$ 35' 37.0" W& D. I. Machado      & 28 cm   & 35 \\
FOZ             &25$^o$ 26' 05.0" S& L. L. Trabuco      & Raptor/Merlin  &    \\
X57             &184 m          &                    & 0.73 "/px &  \\
\hline
Guaratinguet\'a/SP-Brazil&45$^o$ 11' 25.5" W& R. Sfair           & 40 cm   & 24 \\
FEG             &22$^o$ 48' 05.5" S& T. de Santana         & Raptor/Merlin  &    \\
XXX    &543 m          & L. A. Boldrin      & 0.55 "/px &  \\
                &                  & G. Borderes-Mota   &         &    \\
                &                  & T. S. Moura           &         &    \\
                &                  & T. Akemi           &         &    \\
                &                  & B. C. B. Camargo   &         &    \\
                &                  & O. C. Winter          &         &    \\
\hline
Vit\'oria/ES-Brazil      &40$^o$ 19' 00.0" W& M. Malacarne       & 35 cm   & 8 \\
GOA             &20$^o$ 17' 52.0" S& J. O. Miranda      & SBIG/ST-8X-ME    &    \\
XXX    &26 m          & F. Krieger         & 0.65 "/px &  \\
\hline
Curitiba/PR-Brazil     &49$^o$ 11' 45.8" W& F. Braga-Ribas     & 25 cm   & 5 \\
UTF             &25$^o$ 28' 24.6" S& A. Crispim         & Watec/910HX  &    \\
XXX    &861 m           &                    & 0.32 "/px &  \\
\hline
Haute de Province/France &05$^o$ 42' 56.5" E& V. Robert       & 120 cm   & 3 \\
OHP             &43$^o$ 55' 54.7" N& V. Lainey          & Andor/CCD42-40    &    \\
511             &633 m          &                    & 0.38 "/px &  \\

\hline
\hline
\label{tb:obs_sites}
\end{tabular}
\end{center}
\end{table*}

\subsection{DATA PROCESSING} \label{reducao}

The majority of the observations were acquired in FITS format with time registered in the header. The UTF site recorded observations with a video camera with the time stamped in each frame. The conversion from AVI to FITS and the time extraction were made with the AudeLA\footnote{Website: \url{http://audela.org/}} software. The processing of the FITS images were made in three steps.

First, all images were corrected for bias, dark and flat-field using standard IRAF\footnote{Website: \url{http://iraf.noao.edu/}} procedures \citep{BeS1981}.

The second step was the determination of the satellite's (x,y) centres in the images using the PRAIA package \citep{Assafin2011}. This package measures the object's centroid with a two-dimensional circular symmetric Gaussian fit over pixels within one Full-Width Half Maximum (FWHM = seeing) from the centre. The average error of the centroid measurement was 1/20 of a pixel. Using the nominal pixel scale of the instruments, this translates to errors in the range 16 to 36 mas.

The third step was the application of the mutual approximation method itself, described in Section \ref{Method}. We fitted the observed and ephemeris distance curves for the determination of auxiliary ephemeris central instants of time, impact parameter and relative velocities. After the corrections for solar phase angle, atmospheric refraction, diurnal and annual aberration, we obtained the final observed central instants, as well as the observed impact parameters and relative velocities, and their errors. It turned out that a fourth-degree polynomial was used to fit all the distance curves. A Python-based software \citep{Astropy2013} was specially developed for performing all the computations of this step. 

Thus, at the end of the data processing, we obtained the central instant of the maximum apparent approximation ($t_0$) between both satellites, their impact parameter ($d_0$), their relative velocity ($v_0$) at $t_0$, also in the sky plane, and the errors for all these parameters. Without any scaling, $d_0$ and $v_0$ (and their errors) are measured in pixels and pixel per second.

\section{RESULTS}\label{res_APPROX}

The 2016-2018 observational campaign reported in this paper started in February 2016 and ended in August 2018. A total of 66 events were successfully observed. For 28 mutual approximations, simultaneous observations were made at two or more sites. In total, 104 independent observations were obtained.

\begin{figure*}
\begin{centering}
\subfigure{\includegraphics[height=06cm]{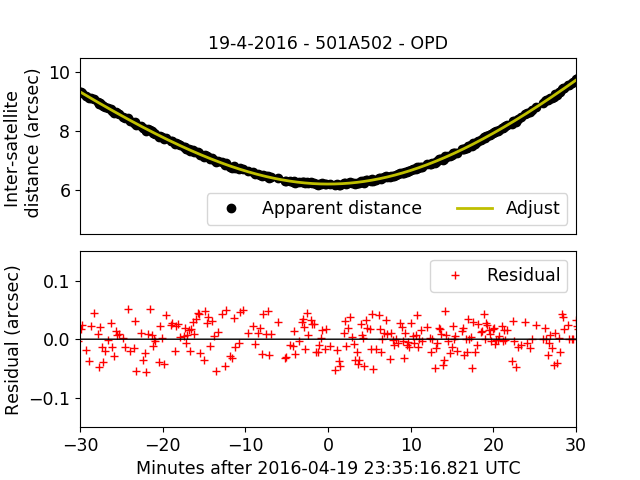}\label{Fig:exemplo_A}}
\subfigure{\includegraphics[height=06cm]{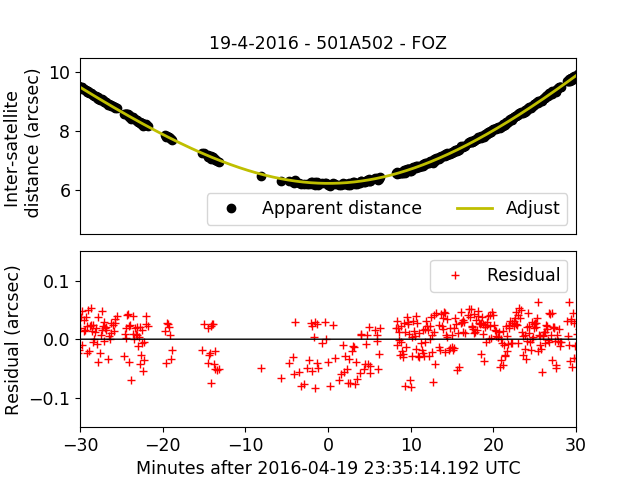}\label{Fig:exemplo_B}}
\subfigure{\includegraphics[height=06cm]{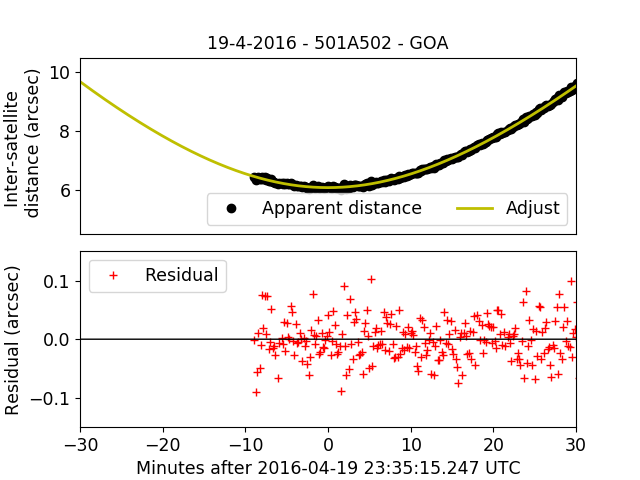}\label{Fig:exemplo_C}}
\subfigure{\includegraphics[height=06cm]{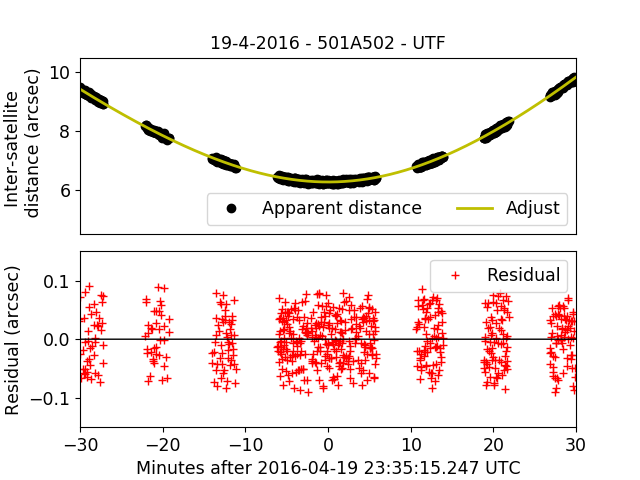}\label{Fig:exemplo_D}}
\subfigure{\includegraphics[height=06cm]{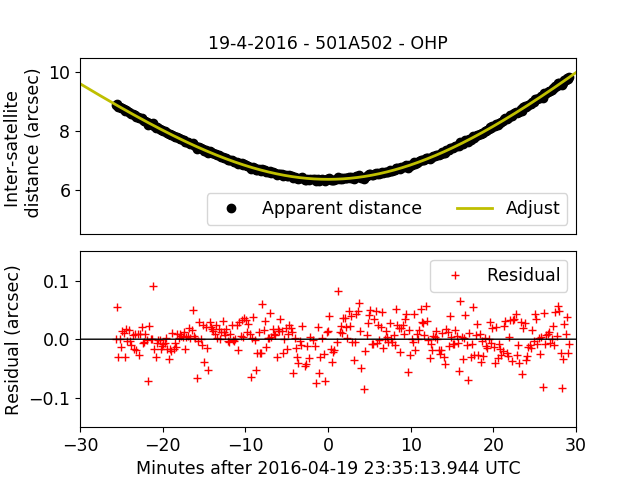}\label{Fig:exemplo_E}}
\caption{Observed apparent sky plane distance curves between Io and Europa in the mutual approximation of April, $19^{th}$ 2016 for 5 sites -- OPD, FOZ, GOA, UTF and OHP. Inter-satellite distances, $d$, are the black dots and the model fitted is the yellow line. In the bottom, the red crosses are the residual of the fitting. We used the nominal pixel scale (Table \ref{tb:obs_sites}) for each site to convert the apparent distances from pixels to arcseconds.}
\label{fig:approx_examples}
\end{centering}
\end{figure*}

The multiple coverage observational strategy reduced the number of events lost by bad weather or instrumental issues. An extreme example was the event between Io and Europa on April, 19$^{th}$ 2016. This approximation was observed by five sites: OPD, FOZ, GOA, UTF and OHP. Fig. \ref{fig:approx_examples} contains the distance curves obtained by each observatory. For comparison, we used the nominal pixel scale (Table \ref{tb:obs_sites}) for each site to transform the apparent sky plane distance from pixels to arcseconds. The differences between the observations and the ephemeris \emph{jup310} from JPL\footnote{Website: \url{http://www.jpl.nasa.gov/}} were -0.2, -4.8, -7.8, -0.5 and -5.5 mas and the precision 3.8, 8.2, 8.2, 12.0 and 5.6 mas, respectively. These observations combined represent an offset of -3.8 mas with a standard deviation of 2.9 mas.
 
The results of this campaign can be found on Tables \ref{tb:results_16} and \ref{tb:results_17}, which contain the date of the event and the satellites' pairs in the form "$S_{i} A S_{j}$", where 501 stands for Io, 502 for Europa, 503 for Ganymede and 504 for Callisto. We furnish the sites involved on each observation (using the alias defined in Table \ref{tb:obs_sites}). For each site, we give the obtained central instant ($t_0$) and its uncertainties ($\sigma t_0$) in seconds of time and in mas, respectively, and the difference between the observed central instant and that determined by using the (topocentric) ephemeris \emph{jup310} with DE435 ($\Delta t_0$), in seconds of time and in mas. All times are UTC. In the last column, we have the label $N$ of each mutual approximation, which is a sequential number following the chronological order of the events. Table \ref{tb:results_16} displays the results of the 48 distance curves obtained in 2016 and Table \ref{tb:results_17} shows the results of the 25 curves observed in 2017 and 31 obtained in 2018.   

\begin{table*}
\begin{center}
\caption{Central Instant -- 2016 Mutual approximations.}
\begin{tabular}{ccccrrrrc}
\hline
\hline
Date &   Event & Observer   &      $t_0$ UTC  &   $\sigma t_0$ & $\sigma t_0$ &  $\Delta t_0$  &  $\Delta t_0$& N\\
(dd-mm-yy)  & & &  (hh:mm:ss.ss)  &   (s)   &  (mas) &  (s)  &   (mas) &\\ 
\hline
\hline
03-2-2016  & 502A503 & OPD & 04:48:01.1&  4.2 & 30.4 & +0.9 & + 6.3 & 1 \\
08-2-2016  & 501A502 & FOZ & 06:29:38.4&  0.6 &  2.6 & +1.5 & + 6.8 & 2  \\
15-2-2016  & 502A503 & FOZ & 08:39:28.5&  1.1 &  4.8 & -0.7 & - 2.8 & 3  \\
24-2-2016  & 501A503 & OPD & 01:53:25.5&  1.1 &  7.1 & -0.2 & - 1.6 & 4  \\
           &         & FEG & 01:53:27.3&  4.0 & 24.7 & +1.5 & + 9.6 & 4  \\
25-2-2016  & 501A503 & GOA & 23:55:58.2&  2.4 &  8.8 & -1.6 & - 6.0 & 5  \\
04-3-2016  & 501A502 & GOA & 02:09:59.3&  2.3 &  7.7 & +1.8 & + 6.1 & 6  \\
18-3-2016  & 501A502 & OPD & 06:53:17.0&  2.5 &  5.9 & +9.7 & +22.4 & 7  \\
02-4-2016  & 501A502 & OPD & 05:46:03.2&  2.5 &  7.0 & -3.2 & - 9.2 & 8  \\
           &         & FOZ & 05:45:57.1&  2.2 &  6.4 & -9.3 & -26.7 & 8  \\
           &         & FEG & 05:45:59.1&  3.8 & 10.8 & -7.4 & -21.1 & 8  \\
02-4-2016  & 501A504 & OPD & 23:24:20.4&  1.2 &  6.6 & -8.1 & -44.9 & 9  \\
           &         & FOZ & 23:24:22.4&  1.4 &  7.5 & -6.2 & -33.9 & 9  \\
           &         & FEG & 23:24:22.3&  3.5 & 19.1 & -6.3 & -34.5 & 9  \\
12-4-2016  & 501A504 & OPD & 04:35:29.7&  8.9 & 51.6 & +5.2 & +30.3 & 10  \\
           &         & FOZ & 04:35:31.1&  1.1 &  6.4 & +6.6 & +38.3 & 10  \\
           &         & FEG & 04:35:29.1&  2.5 & 14.5 & +4.7 & +27.1 & 10  \\
12-4-2016  & 501A502 & FOZ & 04:45:49.0& 10.1 & 10.5 &+19.1 & +19.8 & 11  \\
12-4-2016  & 502A504 & FOZ & 05:01:34.6&  1.9 & 11.2 & +0.0 & + 0.1 & 12  \\
           &         & FEG & 05:01:36.1&  4.2 & 25.2 & +1.6 & + 9.8 & 12  \\
12-4-2016  & 502A504 & OPD & 21:17:16.2&  0.8 &  2.9 & -7.2 & -25.1 & 13  \\
19-4-2016  & 501A502 & OPD & 23:35:15.3&  1.0 &  3.8 & -0.1 & - 0.2 & 14  \\
           &         & FOZ & 23:35:14.2&  2.1 &  8.2 & -1.3 & - 4.8 & 14  \\
           &         & GOA & 23:35:13.3&  2.2 &  8.2 & -2.1 & - 7.8 & 14  \\
           &         & UTF & 23:35:15.2&  3.2 & 12.0 & -0.2 & - 0.6 & 14  \\
           &         & OHP & 23:35:13.9&  1.5 &  5.6 & -1.4 & - 5.5 & 14  \\
20-4-2016  & 501A502 & OHP & 20:15:57.8&  1.8 &  8.7 & -3.6 & -17.0 & 15  \\
24-4-2016  & 502A504 & OPD & 22:35:12.0&  0.5 &  3.7 & -1.5 & -11.6 & 16  \\
           &         & UTF & 22:35:13.1&  2.6 & 19.6 & -0.4 & - 3.3 & 16  \\
29-4-2016  & 501A503 & OPD & 00:32:28.1&  2.4 & 16.2 & -1.4 & - 9.2 & 17  \\
           &         & UTF & 00:32:28.6&  4.2 & 28.0 & -0.9 & - 5.8 & 17  \\
02-5-2016  & 501A503 & OPD & 01:08:50.3&  1.5 & 10.5 & +0.5 & + 3.3 & 18  \\
           &         & FOZ & 01:08:50.7&  2.3 & 16.7 & +0.8 & + 6.0 & 18  \\
           &         & FEG & 01:08:49.1&  1.8 & 12.7 & -0.7 & - 4.8 & 18  \\
           &         & UTF & 01:08:51.1&  4.5 & 32.1 & +1.3 & + 9.2 & 18  \\
03-5-2016  & 501A503 & OPD & 01:04:55.4&  1.3 &  4.2 & +5.4 & +18.2 & 19  \\
           &         & UTF & 01:04:55.5&  1.9 &  6.4 & +5.4 & +18.4 & 19  \\
06-5-2016  & 502A503 & OPD & 00:59:06.8&  6.5 & 31.6 & +3.2 & +15.6 & 20  \\
19-5-2016  & 502A504 & FOZ & 22:52:31.9&  1.0 &  6.6 & -1.4 & - 9.1 & 21  \\
27-5-2016  & 501A503 & FEG & 02:00:21.8&  5.5 & 34.2 & +0.1 & + 0.9 & 22  \\
17-6-2016  & 502A503 & OPD & 00:48:02.9&  1.3 &  9.0 & -0.3 & - 2.2 & 23  \\
           &         & FEG & 00:48:07.0&  4.8 & 34.3 & +3.8 & +26.7 & 23  \\
28-6-2016  & 501A502 & OPD & 23:58:57.1&  1.4 &  6.5 & +0.2 & + 0.8 & 24  \\
           &         & FEG & 23:58:59.0&  1.1 &  5.2 & +2.1 & + 9.7 & 24  \\
29-6-2016  & 501A503 & OPD & 22:36:02.2&  0.5 &  2.9 & +1.2 & + 6.6 & 25  \\
           &         & FEG & 22:36:02.9&  1.2 &  6.7 & +1.8 & +10.3 & 25  \\
08-7-2016  & 501A502 & OPD & 21:51:35.5&  0.6 &  3.2 & +3.0 & +14.4 & 26  \\
           &         & FEG & 21:51:32.6&  3.3 & 16.2 & +0.0 & + 0.2 & 26  \\
\hline
\hline
\label{tb:results_16}
\end{tabular}
\end{center}
\emph{Note}: Results of the mutual approximation campaign for 2016: 501 stands for Io, 502 for Europa, 503 for Ganymede and 504 for Callisto. $\sigma t_0$ is the central instant error in seconds of time and in mas (using the relative velocity in each event obtained with the ephemeris) and $\Delta t$ is the comparison between the observation and the ephemeris \emph{jup310} (with DE435) from JPL in the sense \emph{"observation minus ephemeris"} in seconds of time and in mas. N is a sequential number with time that labels each observed mutual approximation. Time is UTC.
\end{table*}

\begin{table*}
\begin{center}
\caption{Central Instant -- 2017 and 2018 Mutual approximations.}
\begin{tabular}{ccccrrrrc}
\hline
\hline
Date &   Event & Observer   &      $t_0$ UTC  &   $\sigma t_0$ & $\sigma t_0$ &  $\Delta t_0$  &  $\Delta t_0$& N\\
(dd-mm-yy)  & & &  (hh:mm:ss.ss)  &   (s)   &  (mas) &  (s)  &   (mas) &\\ 
\hline
\hline
07-2-2017  & 502A503 & FOZ & 04:36:54.1&  1.0 &  7.5 & +2.0 & +14.4 & 27  \\
26-2-2017  & 502A503 & FOZ & 04:32:43.5&  1.3 &  9.0 & +2.6 & +17.7 & 28  \\
27-2-2017  & 501A502 & FOZ & 03:36:51.3&  1.1 &  5.7 & +3.6 & +18.8 & 29  \\
07-3-2017  & 501A502 & FOZ & 03:00:44.4& 32.9 & 31.0 & -2.2 & - 2.1 & 30  \\
14-3-2017  & 501A503 & FOZ & 07:19:33.8&  1.1 &  3.4 & +0.9 & + 2.8 & 31  \\
04-4-2017  & 501A503 & OHP & 20:43:34.4&  0.7 &  6.3 & +1.9 & +17.9 & 32  \\
06-4-2017  & 501A503 & FEG & 03:46:43.1&  2.2 & 13.1 & +3.9 & +23.0 & 33  \\
08-4-2017  & 501A502 & FOZ & 01:52:40.5&  1.0 &  7.6 & +2.3 & +16.9 & 34  \\
13-4-2017  & 501A502 & FOZ & 05:49:28.3&  1.0 &  5.8 & +1.8 & +10.0 & 35  \\
06-5-2017  & 502A503 & GOA & 02:16:30.2&  1.7 & 12.1 & +2.9 & +20.9 & 36  \\
08-5-2017  & 501A502 & FOZ & 01:11:26.5&  1.0 &  4.5 & +1.2 & + 5.5 & 37  \\
13-5-2017  & 501A503 & FOZ & 04:47:32.1&  1.0 &  7.0 & +1.0 & + 6.7 & 38  \\
15-5-2017  & 501A502 & FEG & 03:23:43.1&  1.7 &  6.8 & +2.0 & + 8.4 & 39  \\
31-5-2017  & 501A503 & FEG & 22:30:36.2& 27.9 & 14.7 & +4.0 & + 2.1 & 40  \\
08-6-2017  & 501A502 & GOA & 23:48:58.1&  1.8 &  4.3 & +2.9 & + 6.9 & 41  \\
           &         & FEG & 23:48:57.1&  7.5 & 17.6 & +1.9 & + 4.4 & 41  \\
23-6-2017  & 501A502 & FOZ & 23:17:09.0&  1.1 &  2.8 & +2.4 & + 5.7 & 42  \\
           &         & GOA & 23:17:07.7&  1.9 &  4.6 & +1.2 & + 2.9 & 42  \\
06-7-2017  & 501A502 & FOZ & 22:58:42.6&  1.4 &  3.4 & +1.0 & + 2.4 & 43  \\
           &         & FEG & 22:58:41.1& 19.4 & 48.3 & -0.5 & - 1.3 & 43  \\
25-7-2017  & 502A503 & FOZ & 22:40:24.8&  1.2 &  4.9 & +2.9 & +11.6 & 44  \\
           &         & FEG & 22:40:21.3&  3.3 & 13.1 & -0.5 & - 1.9 & 44  \\
02-8-2017  & 501A502 & FEG & 23:38:20.0&  7.7 & 28.7 & +4.8 & +17.6 & 45  \\
10-8-2017  & 501A502 & FOZ & 23:41:23.6& 48.2 & 30.6 & -1.9 & - 1.2 & 46  \\
24-8-2017  & 503A504 & FEG & 22:35:37.6&  6.6 & 16.1 & +1.4 & + 3.3 & 47  \\
\hline
05-03-2018 & 501A502 & FOZ & 05:10:29.7&  0.6 &  3.7 & -0.4 & - 2.4 & 48 \\
11-03-2018 & 501A503 & OPD & 05:40:46.7&  1.8 &  4.4 & +0.3 & + 0.6 & 49 \\
           &         & FOZ & 05:40:47.0&  2.0 &  5.0 & +0.5 & + 1.3 & 49 \\
12-03-2018 & 501A502 & OPD & 07:20:57.6&  0.5 &  3.0 & +0.1 & + 0.4 & 50 \\
           &         & FOZ & 07:20:58.8&  1.4 &  8.4 & +1.2 & + 7.3 & 50 \\
17-03-2018 & 501A502 & FOZ & 03:15:03.2&  0.8 &  7.0 & +0.2 & + 1.5 & 51 \\
17-03-2018 & 502A504 & FOZ & 03:41:06.1&  2.1 & 13.0 & +1.8 & +11.1 & 52 \\
24-03-2018 & 501A502 & FOZ & 05:18:47.9&  0.7 &  5.7 & +1.4 & +11.3 & 53 \\
06-04-2018 & 501A502 & OPD & 02:40:32.0&  1.2 &  8.8 & +1.1 & + 8.2 & 54 \\
           &         & FOZ & 02:40:31.4&  1.0 &  7.7 & +1.2 & + 8.5 & 54 \\
11-06-2018 & 502A503 & FEG & 23:03:46.0&  1.8 & 12.4 & -0.4 & - 3.0 & 55  \\
           &         & GOA & 23:03:45.1&  1.2 &  8.3 & -1.3 & - 9.2 & 55 \\
19-06-2018 & 502A503 & FOZ & 01:55:19.9&  1.1 &  7.6 & +0.3 & + 2.4 & 56 \\
22-06-2018 & 501A503 & OPD & 02:17:12.6&  4.5 &  5.7 & -2.0 & - 2.5 & 57 \\
           &         & FOZ & 02:17:12.5&  5.6 &  7.0 & -4.7 & - 6.0 & 57 \\
           &         & FEG & 02:17:09.5&  7.2 &  9.0 & -5.0 & - 6.3 & 57  \\
           &         & GOA & 02:17:09.9&  6.5 &  8.2 & -2.0 & - 2.5 & 57 \\
23-06-2018 & 501A502 & FOZ & 00:40:47.4&  1.1 &  9.1 & -1.9 & -16.1 & 58 \\
07-07-2018 & 501A503 & OPD & 00:30:56.8&  1.1 &  6.3 & -0.1 & - 0.8 & 59 \\
           &         & FEG & 00:30:57.0&  2.2 & 12.5 & +0.1 & + 0.4 & 59  \\
11-07-2018 & 502A504 & OPD & 22:48:02.8&  1.4 &  6.7 & -0.4 & - 1.5 & 60  \\
12-07-2018 & 501A504 & OPD & 00:30:30.1&  2.5 &  6.7 & -1.2 & - 3.2 & 61  \\
12-07-2018 & 501A502 & OPD & 01:07:37.4&  1.0 &  5.2 & -0.5 & - 2.5 & 62  \\
           &         & FEG & 01:07:36.3&  2.5 & 12.8 & -1.6 & - 8.1 & 62  \\
13-07-2018 & 502A503 & OPD & 02:01:30.9&  1.1 &  4.4 & +0.1 & + 0.4 & 63  \\
           &         & FEG & 02:01:29.9&  5.4 & 20.9 & -0.9 & - 3.4 & 63  \\
19-07-2018 & 501A504 & OPD & 01:52:08.6&  1.9 &  8.9 & -3.4 & -16.2 & 64  \\
           &         & FOZ & 01:52:09.3&  2.1 & 10.1 & -2.8 & -13.4 & 64  \\
07-08-2018 & 502A503 & OPD & 23:15:18.8&  1.3 &  8.1 & -1.0 & - 6.6 & 65  \\
12-08-2018 & 501A502 & OPD & 23:54:58.4&  1.1 &  3.4 & -1.2 & - 3.5 & 66  \\
           &         & FOZ & 23:54:58.5&  1.2 &  3.5 & -1.1 & - 3.3 & 66  \\
\hline
\hline
\label{tb:results_17}
\end{tabular}
\end{center}
\emph{Note}: Results of the mutual approximation campaign for 2017 and 2018. See note in Table \ref{tb:results_16}.
\end{table*}

\begin{figure*}
\includegraphics[height=06.5cm]{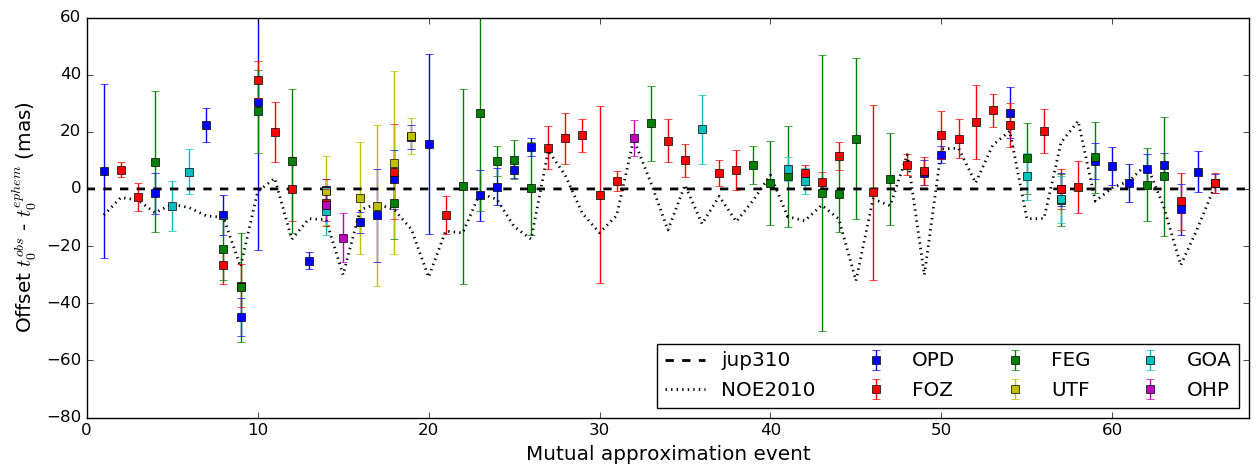}
\caption{Illustration of the APPROX campaign's results. The y-axis is the central instant offset relative to the \emph{jup310} ephemeris with DE435 (dashed line at zero offset) and the error bars represent the error of each observation. The x-axis is the event's label; each colour represents one site. The dotted line represents the difference between the NOE-5-2010-GAL.a and \emph{jup310} ephemeris with DE435.}
\label{Fig:resul_approx}
\end{figure*}

Our results are also illustrated in Fig. \ref{Fig:resul_approx} which displays central instant offsets in mas with respect to the \emph{jup310} ephemeris (dashed line at zero offset) for each mutual approximation. The different colours represent different sites and the dotted line is the difference between NOE-5-2010-GAl.a and \emph{jup310} ephemeris. The RMS between our observations and the \emph{jup310} and NOE-5-2010-GAl.a ephemeris were 14.4 and 18.2 mas, respectively. 

\begin{figure}
\begin{centering}
\includegraphics[height=06cm]{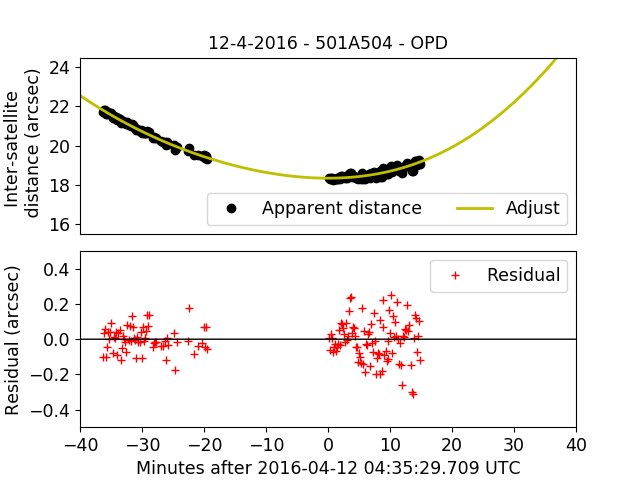}
\caption{Observed apparent sky plane distance curves for Io and Callisto in the mutual approximation of April, $12^{th}$ 2016 seen from OPD. Inter-satellite distances, $d$, are the black dots and the model fitted is the yellow line. In the bottom, the red crosses are the residual of the fitting. We used the nominal pixel scale (Table \ref{tb:obs_sites}) to convert the apparent distances from pixels to arcseconds. The internal error for this observation was large, 8.9 seconds of time (51.6 mas), due to the central gap and to the lack of observations 15 minutes after the central instant.}
\label{fig:bad}
\end{centering}
\end{figure}

According to Tables \ref{tb:results_16} and \ref{tb:results_17}, very few observations had central instants with internal errors worse than the ideal 30 mas suggested in \cite{Lainey2009} and \cite{Lainey2016} for an effective contribution to the study of tidal forces in the Jovian system - 9 observations out of 104 (about 9\%). About 87\% (90 observations) had uncertainties below 20 mas and 65\% (67 observations) below 10 mas. One extreme example was the mutual approximation between Io and Callisto observed at OPD in April, 12$^{th}$  2016 (N = 10). The internal error was 8.9 seconds (51.6 mas). This event was heavily affected by bad weather, presenting a 20 minutes gap before the central instant, with observations having to stop just 15 minutes after the central instant, see Fig. \ref{fig:bad}. All observations with internal errors worse than about 20 mas were affected at some extent by bad weather conditions and/or instrumental issues, such as gaps in the curve or low signal to noise ratio (SNR). These scenarios were predicted and explored in our simulations in Section \ref{simulacao}.

\section{EPHEMERIS FITTING PROCEDURE}\label{ajuste}

In order to create an ephemeris, it is necessary to fit a dynamical model to observations. In the case of natural satellites, the fitting is made by using the standard method of variational equations (see for example \citealt{Lainey2004a, Lainey2004b}). Here, we present for ephemeris developers a method that permits to add central instants from mutual approximations to ephemeris fitting, by the development of more adequate conditional equations to the problem.

In the case of mutual approximations, we should in principle solve for the partial derivatives $\dfrac{\partial t_0}{\partial c_l}$ to obtain the conditional equations \eqref{Eq:1} of the problem:
\begin{equation}
t_0^o - t_0^c = \Delta t_0 = \sum \dfrac{\partial t_0^c}{\partial c_l} \Delta c_l ~, \label{Eq:1}
\end{equation}
where $c_l$ represents each of the $l$ parameters that we are fitting, usually the initial positions and velocities ($X_0$,$Y_0$,$Z_0$,$\dot{X}_0$,$\dot{Y}_0$,$\dot{Z}_0$) for each body in the integration, and other parameters such as the masses, $J_{2}$, $J_{4}$, etc. $\Delta c_l$ represents the correction for each fitted parameter. $t_0^c$ is the central instant computed by the dynamical model, $t_0^o$ is the central instant obtained from the observations and the difference $t_0^o - t_0^c = \Delta t_0$ represents the "observed minus computed" offset.
  
However, equations \eqref{Eq:1} cannot be solved analytically and a numerical approach consumes too much CPU time (see \citealt{Emelyanov2017}). Fortunately, we can develop equivalent equations to the problem which are solvable.

Consider the apparent distance in the sky plane $s_{ij}$ between two satellites $i$ and $j$. $s_{ij}$ is minimum at the central instant $t_0$, i.e. $\dfrac{ds^o}{dt}(t_0) = 0$. Knowing that, we can write the equation \eqref{Eq:Var1}:
\begin{equation}
\dfrac{ds^o}{dt}(t_0) - \dfrac{ds^c}{d t}(t_0) = \sum \dfrac{\partial}{\partial c_l} \left( \dfrac{ds^c}{dt}(t_0) \right) \Delta c_l ~\label{Eq:Var1}
\end{equation}
where $\dfrac{ds^c}{d t}(t_0)$ is the value computed by the dynamical model, and the difference $\dfrac{ds^o}{dt}(t_0) - \dfrac{ds^c}{dt}(t_0)$ also represents an "observed minus computed" offset. Equation \eqref{Eq:Var1} is a more suitable conditional equation to the problem. It can be rewritten as follows.

The apparent distance $s_{ij}$ between satellites $i$ and $j$ can be written as in equation \eqref{Eq:s}:

\begin{equation}
s_{ij} = \sqrt{\Delta x^2_{ij} + \Delta y^2_{ij}} \label{Eq:s},
\end{equation}
with
    
\begin{eqnarray}
\Delta x_{ij} &\simeq& (\alpha_i - \alpha_j)\cos(\delta_m) \label{Eq:4} ~,\\
\Delta y_{ij} &\simeq& \delta_i - \delta_j ~, \label{Eq:5}\\
\delta_m &=& \dfrac{\delta_i + \delta_j}{2} ~, \label{Eq:6}
\end{eqnarray}

where $\alpha_i$ and $\alpha_j$ are the satellites' right ascensions,
$\delta_i$ and $\delta_j$ their declinations and $\delta_m$ is the mean declination of both satellites, this is the first order polynomial approximation for gnomonic projection. The time derivative of the equations \eqref{Eq:4}, \eqref{Eq:5} and \eqref{Eq:6} are:

\begin{eqnarray}
\Delta \dot{x}_{ij} &\simeq& (\dot{\alpha}_i - \dot{\alpha}_j)\cos(\delta_m) - (\alpha_i - \alpha_j)\sin(\delta_m)\dot{\delta}_m ~,\label{Eq:7}\\
\Delta \dot{y}_{ij} &\simeq& \dot{\delta}_i - \dot{\delta}_j ~, \label{Eq:8}\\
\dot{\delta}_m &=& \dfrac{\dot{\delta}_i + \dot{\delta}_j}{2} ~. \label{Eq:9}
\end{eqnarray}

By deriving equation \eqref{Eq:s} over time, we get equation \eqref{Eq:10}:

\begin{equation}
\dfrac{ds_{ij}}{dt} = \dfrac{\Delta x_{ij}\Delta \dot{x}_{ij} + \Delta y_{ij}\Delta \dot{y}_{ij}}{s_{ij}} ~.
\label{Eq:10}
\end{equation}  

Then, we derive equation \eqref{Eq:10} over the $c_l$ to get equation \eqref{Eq:Var2}:

\begin{eqnarray}
\dfrac{\partial}{\partial c_l} \left(\dfrac{ds_{ij}^c}{dt}(t_0) \right) = \dfrac{1}{s_{ij}}&\times& \left[ \Delta\dot{x}_{ij}\dfrac{\partial \Delta x_{ij}}{\partial c_l} + \Delta x_{ij}\dfrac{\partial \Delta\dot{x}_{ij}}{\partial c_l} \right.\nonumber\\ &+& \left.\Delta\dot{y}_{ij}\dfrac{\partial \Delta y_{ij}}{\partial c_l} + \Delta y_{ij}\dfrac{\partial \Delta\dot{y}_{ij}}{\partial c_l}\right] \nonumber \\ -\dfrac{1}{s_{ij}^3} &\times&\left[ \Delta x_{ij}\Delta \dot{x}_{ij} + \Delta y_{ij}\Delta \dot{y}_{ij}\right] \label{Eq:Var2} \\
&\times& \left[ \Delta x_{ij}\dfrac{\partial \Delta x_{ij}}{\partial c_l} +\Delta y_{ij}\dfrac{\partial \Delta y_{ij}}{\partial c_l}\right]\nonumber  ~. 
\end{eqnarray}

Therefore, we can write the conditional equation \eqref{Eq:Var1} with the use of the explicit form of equation \eqref{Eq:Var2}. 

A similar method was already tested by \cite{Emelyanov2017} and proved to be very efficient when other observations of different kinds (right ascension, declination, relative distance) are fitted together with the central instants from mutual approximations.

\section{Conclusions}\label{conclusao}

The National Observatory (ON) from Brazil, with the collaboration of the IMCCE (Paris Observatory -- France) and Valongo Observatory (UFRJ -- Brazil), organized the APPROX - the mutual approximation campaign for the Galilean Moons. This campaign had the participation of 6 observational sites and obtained 104 distance curves for 66 events. The central instants obtained had an average internal error of 11.3 mas. The external comparisons gave a RMS of 14.4 mas with respect to the JPL \emph{jup310} ephemeris and 18.1 mas with the IMCCE NOE-5-2010-GAL.a ephemeris, using the DE435. About 65\% of our results had precision better than 10 mas, 87\% better than 20 mas and 91\% better than 30 mas. Improvements in the study of the tidal force in the Jovian system requires positions with a precision better than 30 mas \citep{Lainey2009,Lainey2016}.

We used the methane narrow band filter centred at 889 nm with 15 nm width to reduce Jupiter's scattered light. We remark that the time recorded in the images was carefully corrected by the use of GPS receivers or time calibration software. 

The results show that the method of mutual approximations is suitable for small telescopes. They can be used to continually furnish high precision central instants between two satellites.

We also presented a way to fit the observed central instants into dynamical models in order to develop new ephemeris.

The technique of mutual approximations is an alternative high precision astrometric method that serves to improve the orbits of natural satellites. Unlike mutual phenomena, mutual approximations can be observed at any time independently of the equinox of the host planet. Observational campaigns, such as the one presented here, can increase the ephemeris' accuracy and precision and be helpful to space missions aiming at the Jovian system, like the ESA mission JUICE \footnote{Website: \url{http://sci.esa.int/juice/}} and NASA mission Europa Clipper \footnote{Website: \url{https://www.nasa.gov/europa}}.

\section*{Acknowledgments}
This study was financed in part by the Coordena\c{c}\~ao de Aperfei\c{c}oamento de Pessoal de N\'ivel Superior - Brasil (CAPES) - Finance Code 001. Part of this research is suported by INCT do e-Universo, Brazil (CNPQ grants 465376/2014-2). Based in part on observations made at the Laborat\'orio Nacional de Astrof\'isica (LNA), Itajub\'a-MG, Brazil and at Observatoire de Haute Provence (CNRS), France. BM thanks the CAPES/Cofecub-394/2016-05 grant. RVM acknowledges the following grants: CNPq-306885/2013, CAPES/Cofecub-2506/2015, FAPERJ/PAPDRJ-45/2013 and  FAPERJ/CNE/05-2015. MA thanks to the CNPq (Grants 473002/2013-2 and 308721/2011-0) and FAPERJ (Grant E-26/111.488/2013). J.I.B.C. acknowledges CNPq grant 308150/2016-3. RS e OCW acknowledges Fapesp proc. 2016/24561-0, CNPq proc 312813/2013-9 and 305737/2015-5. FBR acknowledges CNPq support, process 309578/2017-5. VL's research was supported by an appointment to the NASA Postdoctoral Program at the NASA Jet Propulsion Laboratory, California Institute of Technology, administered by Universities Space Research Association under contract with NASA.

%%%%%%%%%%%%%%%%%%%%%%%%%%%%%%%%%%%%%%%%%%%%%%%%%%

%%%%%%%%%%%%%%%%% APPENDICES %%%%%%%%%%%%%%%%%%%%%

%\appendix



%%%%%%%%%%%%%%%%%%%%%%%%%%%%%%%%%%%%%%%%%%%%%%%%%%

% Don't change these lines
\bsp	% typesetting comment
\label{lastpage}

\begin{thebibliography}{99}

\bibitem[\protect\citeauthoryear{Arlot et al.}{1982}]{Arlot1982} Arlot J.-E., Bernard A., Bouchet P. et al., 1982,
A\&A, 111, 151
%Primeiro aproximacoes

%\bibitem[\protect\citeauthoryear{Arlot \& Emelyanov}{2009}]{AeE2009} Arlot J.-E. \& Emelyanov N. V., 2009,
%A\&A, 503, 631
%ARLOT E EMELYANOV NSDB

\bibitem[\protect\citeauthoryear{Arlot et al.}{2012}]{Arlot2012} Arlot J.-E., Emelyanov N. V., Lainey V. et al., 2012,
A\&A, 544, A29
%PHEMU SATURNO

\bibitem[\protect\citeauthoryear{Arlot et al.}{2013}]{Arlot2013} Arlot J.-E., Emelyanov N. V., Aslan Z. et al., 2013,
A\&A, 557, A4
%PHEMU URANO

\bibitem[\protect\citeauthoryear{Arlot et al.}{2014}]{Arlot2014} Arlot J.-E., Emelyanov N. V., Varfolomeev M. I. et al., 2014,
A\&A, 572, A120
%ARLOT PHEMU 2009

\bibitem[\protect\citeauthoryear{Assafin et al.}{2011}]{Assafin2011} Assafin M. et al., 2011,
in Gaia follow-up network for the solar system objects :Gaia FUN-SSO
workshop proceedings, held at IMCCE -Paris Observatory, France,
November 29 - December 1, 2010. ISBN 2-910015-63-7, ed. P. Tanga
\& W. Thuillot, 85-88
%PRAIA

\bibitem[\protect\citeauthoryear{Astropy Collaboration et al}{2013}]{Astropy2013} Astropy Collaboration, Robitaille, T. P., Tollerud, E. J., et al., 2013, A\& A, 558, A33
%ASTROPY

\bibitem[\protect\citeauthoryear{Butcher \& Stevens}{1981}]{BeS1981} Butcher E. \& Stevens R., 1981,
News Letter Kitt Peak National Observatory, 16, 6
%BIAS-FLAT-IRAF

\bibitem[\protect\citeauthoryear{Charnoz et al.}{2011}]{Charnoz2011} Charnoz S. et al., 2011,
Icarus, 216, 535
%Formacao de satelites

\bibitem[\protect\citeauthoryear{Crida \& Charnoz}{2012}]{Crida_2012} Crida A. \& Charnoz S., 2012,
Science, 338, 1196
%Formacao dos satelites

\bibitem[\protect\citeauthoryear{De Sitter}{1928}]{Sitter1928} De Sitter W., 1928,
Leiden Ann., 16, 1
%OLD AS LAINEY 2009

\bibitem[\protect\citeauthoryear{Dias-Oliveira et al.}{2013}]{Diasoliveira2013} Dias-Oliveira A., Vieira-Martins R., Assafin M., et al., 2013,
MNRAS, 432, 225
%ALEX PHEMU 2009

\bibitem[\protect\citeauthoryear{Emelyanov}{2009}]{Emelyanov2009} Emelyanov N. V., 2009,
MNRAS, 394, 1037
%PHEMU 2002-2003

\bibitem[\protect\citeauthoryear{Emelyanov}{2017}]{Emelyanov2017} Emelyanov N. V., 2017,
MNRAS, 469, 4889
%ANALISE APPROX

%\bibitem[\protect\citeauthoryear{Kaasalainen \& Tanga}{2012}]{KeT2004} Kaasalainen M. \& Tanga P., 2004,
%A\&A, 416, 367
%FASE

\bibitem[\protect\citeauthoryear{Karkoschka}{1994}]{Karkoschka1994} Karkoschka E., 1994,
ICARUS, 111, 174
%FILTRO METANO

\bibitem[\protect\citeauthoryear{Karkoschka}{1998}]{Karkoschka1998} Karkoschka E., 1998,
ICARUS, 133, 134
%FILTRO METANO

\bibitem[\protect\citeauthoryear{Kiseleva et al.}{2008}]{Kiseleva2008} Kiseleva T. P., Izmailov I. S., Kiselev A. A. et al., 2008,
Planet. Space Sci., 56, 1908
%ASTROMETRIA CCD CLASSICA

\bibitem[\protect\citeauthoryear{Lainey et al.}{2004a}]{Lainey2004a} Lainey V., Duriez L. \& Vienne A., 2004a,
A\&A, 420, 1171
%MODELO ORBITAL a

\bibitem[\protect\citeauthoryear{Lainey et al.}{2004b}]{Lainey2004b} Lainey V., Arlot J.-E. \& Vienne A., 2004b,
A\&A, 427, 371
%MODELO ORBITAL b

\bibitem[\protect\citeauthoryear{Lainey et al.}{2009}]{Lainey2009} Lainey V., Arlot J. -E., Karatekin O. \& Van Van Hoolst T., 2009,
Nature, 459, 957
%MARE EM IO

\bibitem[\protect\citeauthoryear{Lainey}{2016}]{Lainey2016} Lainey V., 2016,
Celest Mech Dyn Astr, 126, 145
%MARE NO SISTEMA SOLAR

\bibitem[\protect\citeauthoryear{Lieske}{1987}]{Lieske1987} Lieske J. H., 1987,
Astron. Astrophys., 176, 146
%OLD AS LAINEY 2009

\bibitem[\protect\citeauthoryear{Lindegren}{1977}]{Lindegren1977} Lindegren L., 1977,
A\&A, 57, 55
%SOLAR PHASE

\bibitem[\protect\citeauthoryear{Morgado et al.}{2016}]{Morgado2016} Morgado B., M. Assafin, R. Vieira-Martins, J. I. B. Camargo, A. Dias-Oliveira \& A. R. Gomes-J\'unior, 2016,
MNRAS, 460, 4086
%Aproximações

\bibitem[\protect\citeauthoryear{Peng et al.}{2008}]{Peng2008} Peng Q. Y., Vienne A., Lainey V. \& Noyelles B., 2008,
Planet. Space Sci., 419, 1977
%PRECISION PREMIUM

\bibitem[\protect\citeauthoryear{Peng et al.}{2012a}]{Peng2012} Peng Q. Y., He F., Lainey V. \& Vienne A., 2012a,
MNRAS, 56, 1807
%ASTROMETRIA PAR DE SATELITE

\bibitem[\protect\citeauthoryear{Peng et al.}{2012b}]{Peng2012b} Peng Q. Y.,  Vienne A., Zhang Q. F., Desmars J., Yang C. Y. \& He H. F., 2012b, The Astronomical Journal, 144, 170
%DISTORCIONS

\bibitem[\protect\citeauthoryear{Saquet et al.}{2018}]{Saquet2018} Saquet E. V. et al., 2018,
MNRAS, 474, 4730
%PHEMU 2017

%\bibitem[\protect\citeauthoryear{Stone}{1996}]{Stone1996} Stone R. C., 1996,
%Publications of the Astronomical Society of the Pacific, 108, 1051
%COR REFRACAO

\end{thebibliography}
\end{document}